# Accurate experimental ($p$, $\rho$, $T$) data of the ($CO_2$ + $O_2$) binary system for the development of models for CCS processes

Daniel Lozano-Martín[1], Gerald U. Akubue[1,2], Alejandro Moreau[1], Dirk Tuma[3], and César R. Chamorro[1].

[1] Grupo de Termodinámica y Calibración (TERMOCAL), Research Institute on Bioeconomy, Escuela de Ingenierías Industriales, Universidad de Valladolid, Paseo del Cauce, 59, E-47011 Valladolid, Spain.

[2] National Centre for Energy Research and Development & Mechanical Engineering Department, University of Nigeria, Nsukka, 410001 Enugu State, Nigeria

[3] BAM Bundesanstalt für Materialforschung und -prüfung, D-12200 Berlin, Germany.

**Abstract**

The limited availability of accurate experimental data in wide ranges of pressure, temperature, and composition is the main constraining factor for the proper development and assessment of thermodynamic models and equations of state. In the particular case of carbon capture and storage (CCS) processes, there is a clear need for data sets related to the (carbon dioxide + oxygen) mixtures that this work aims to address. This work provides new experimental ($p$, $\rho$, $T$) data for three binary ($CO_2$ + $O_2$) mixtures with mole fractions of oxygen $x(O_2)$ = (0.05, 0.10, and 0.20) mol·mol$^{-1}$, in the temperature range $T$ = (250 to 375) K and pressure range $p$ = (0.5 to 13) MPa. The measurements were performed with a high-precision single-sinker densimeter with magnetic suspension coupling. The density data were obtained with estimated expanded relative uncertainties of 0.02 % for the highest densities, and up to 0.3 % for the lowest ones.The results were compared to the corresponding results calculated by the current reference equations of state for this kind of mixtures, namely the EOS-CG (combustion gases) and the GERG-2008 equation of state, respectively. The EOS-CG yields better estimations in density than the GERG-2008 equation of state. The results from the EOS-GC model show no systematic temperature dependence. For the GERG-2008 model, however, this criterion is significantly less fulfilled.

* Corresponding author e-mail: cescha@eii.uva.es. Tel.: +34 983423756. Fax: +34 983423363

## 1. Introduction

The design and operation of Carbon Capture and Storage (CCS) processes need reliable thermodynamic models able to accurately describe the behavior of the fluid mixtures of $CO_2$ with other gases [1][2]. Besides $CO_2$, the main components involved in those processes related to CCS technologies are $N_2$, $O_2$, Ar, $H_2O$, $H_2$, CO, $H_2S$, and $SO_2$. An approved thermodynamic model able to describe fluid mixtures of these components over an extended $p,T$-region is the GERG-2008 equation of state [3], which is based on a multi-fluid mixture model and explicit in the reduced Helmholtz free energy. However, this EoS was originally developed for natural gas mixtures and thus does not a priori guarantee a high accuracy for mixtures with a composition far from that of typical natural gas mixtures, such as CCS mixtures with perhaps no methane present in them at all, but with $CO_2$ as the main compound.

Some specific EoS have recently been developed for mixtures involved in CCS processes. Demetriades and Graham [4] proposed a pressure-explicit EoS for mixtures of $CO_2$ with small quantities (impurities) of $N_2$, $O_2$, and $H_2$. The range of validity of this model is for pressures up to 16 MPa and temperatures between 273 K and the critical temperature of $CO_2$. A recent research work [5] has proposed a specific model for the binary mixture ($CO_2$ + CO). Gernert and Span proposed an equation of state for the calculation of thermodynamic properties of humid gases, combustion gases, and $CO_2$-rich mixtures typical in CCS processes, the so-called EOS-CG (Equation of State for Combustion Gases and Combustion Gas-like Mixtures) [6]. This equation of state, with a functional structure similar to the GERG-2008, based on a multi-fluid mixture model explicit in the reduced Helmholtz free energy, has a wider range of validity in temperature and pressure, and more components are considered. The EOS-CG has been developed for 6 constituting pure components: $CO_2$, $N_2$, $O_2$, Ar, $H_2O$, and CO. Unfortunately, binary specific departure functions were developed (or, in some case, taken from the GERG-2008 EoS) for only some of the 15 resulting binary mixtures, those for which sufficiently accurate experimental data were available,. As for the binary mixture (carbon dioxide + oxygen), no departure function has yet been developed due to limited experimental data.

High-accuracy density data are of great relevance for the development of reliable equations of state for CCS processes [7]. In this work, accurate density measurements for three binary mixtures of carbon dioxide with oxygen (nominal amount-of-substance fraction $x(O_2)$ = 0.05, 0.10, 0.20) are presented. Measurements were performed at temperatures between (250 and 375) K and pressures up to 13 MPa, using a single-sinker densimeter with magnetic suspension coupling, which is one of the state-of-the-art methods for density determination over wide ranges of temperature and pressure. In order to achieve the highest accuracy in composition, the binary mixtures for this investigation were prepared gravimetrically according to the ISO 6142-1 [8], a method that qualifies for the production of reference materials. The experimental results are compared with the calculations by both the GERG-2008 equation of state and the EOS-CG, and also with the limited experimental data available in the literature.

## 2. Experimental

### 2.1. Mixture preparation

Three ($CO_2$ + $O_2$) binary mixtures were prepared at the Federal Institute for Materials Research and Testing (Bundesanstalt für Materialforschung und -prüfung, BAM) in Berlin, Germany, following the recommendations given in the standard ISO 6142-1 [8].

Purity, supplier, molar mass, and critical parameters of the pure compounds (obtained from the reference equations of state for carbon dioxide [9] and oxygen [10]) are given in Table 1. Table 2 shows the gravimetric composition and its corresponding absolute expanded uncertainty ($k$ = 2) of the three mixtures. Carbon dioxide and oxygen were used without further purification, but information on impurities from the specification was considered in the preparation of the mixture.

Prior to the preparation, the mass portions of the constituting compounds had to be calculated. The phase boundary was calculated applying the GERG-2008 EoS, and the gas portions were adapted in such a way that a homogeneous gas phase would be inside the cylinder at room temperature.

The preparation of the mixtures (BAM reference gas mixture G 033) consisted of two consecutive steps in which the pure compound was transferred. First, pure carbon dioxide was introduced into the evacuated recipient cylinder. The carbon dioxide was taken from the liquid phase, which required an upside-down mounting of the feed cylinder. Heating of the feed cylinder was administered during the transfer to enlarge the pressure difference between the two cylinders. The oxygen was introduced in the second step from a feed cylinder that had sufficient internal pressure to ensure the transfer of the required mass portion. During each filling step, the recipient cylinder stood on the platform of an electronic comparator balance (Sartorius

LA 34000P-0CE, Sartorius AG, Göttingen, Germany, weighing range: 34 kg, readability: 0.1 g) to monitor the feed stream. The exact mass of the gas added was determined after each filling step using a high-precision mechanical balance (Voland model HCE 25, Voland Corp., New Rochelle NY, USA, weighing range: 25 kg, readability: 2.5 mg).

The following gas portions were determined that resulted in the final pressures:

| | | | |
|---|---|---|---|
| Cylinder 1007-180626 ($x(O_2) = 0.05$): | 1088.762 g $CO_2$ | 41.946 g $O_2$ | $p = 5.8$ MPa |
| Cylinder 1008-180626 ($x(O_2) = 0.10$): | 1137.833 g $CO_2$ | 91.775 g $O_2$ | $p = 6.4$ MPa |
| Cylinder 9085-180116 ($x(O_2) = 0.20$): | 1073.480 g $CO_2$ | 195.014 g $O_2$ | $p = 6.8$ MPa |

After the filling, the three mixtures were homogenized by subsequent heating and rolling for approximately 8 hours each.

Prior to density determination, the cylinders were validated at BAM by gas chromatography (GC) on a multichannel process analyzer (Siemens MAXUM II, Siemens AG, Karlsruhe, Germany). Details of configuration and operation are given in a previous paper [11]. The matrix compound $CO_2$ was not analyzed for these mixtures, as the $CO_2$ content is outside the operational range of the GC. The bracketing method outlined in ISO 12963 was applied for validation [12]. The calibration mixtures were prepared by the gravimetric method using the same procedures as for the research samples. Any interdependence between both calibration gases and the research samples was thus avoided. The results of the GC analysis and the corresponding (gravimetric) composition of the binary mixtures used for validation can be found in Table 3. The deviations between gravimetric realization and GC analyses were low enough to pass the criteria for certification.

## 2.2. Equipment description

The ($p$, $\rho$, $T$) data were measured using a single-sinker magnetic suspension densimeter (SSMSD) especially designed for density measurements of pure gases and gaseous mixtures. Details of the equipment and measurement procedure have been previously described by Chamorro et al. [13], Mondéjar et al. [14], and Lozano-Martín et al. [15]. The measuring method, originally developed by Brachthäuser et al. [16] and improved by Klimeck et al. [17], operates on the Archimedes' principle. A magnetic suspension coupling system allows the determination of the buoyancy force on a sinker immersed in the gas so that accurate density measurements of fluids over wide temperature and pressure ranges can be obtained. The sinker used in this work was cylindrical and made of monocrystalline silicon with a mass of 61.59181 ± 0.00016 g and a volume of 26.444 ± 0.003 $cm^3$ ($k$ = 2), measured at $T$ = 293.05 K and $p$ = 1.01134 bar, and determined at the Mass Division of the Spanish National Metrology Institute (Centro Español de Metrología, CEM).

The magnetic coupling is formed by an electromagnet hanging from the lower hook of an analytical balance (Mettler Toledo XPE205DR, Mettler Toledo GmbH, Gießen, Germany, weighing range: 81 g, readability: 0.01 mg, extended weighing range: 220 g) and a permanent magnet inside the measuring cell. The permanent magnet is fixed to a sinker support, which allows the sinker to be coupled and decoupled from the balance. The magnetic coupling has two different positions: the zero position (ZP) and the measuring position (MP). A load compensation system consists of two calibrated masses, one made of tantalum and the other of titanium, which can be put on the upper pan of the balance. Both masses have approximately the same volume (4.9 $cm^3$) and the mass difference between both is similar to the mass of the sinker. This characteristic system and the differential nature of the measurement procedure allow the balance to operate near its zero point and to perform a periodic calibration of the balance, free from the air buoyancy corrections. The two masses were provided by Rubotherm GmbH, Bochum, Germany, and their individual mass and volume were also accurately determined at the Mass Division of the CEM.

The temperature inside the measuring cell is determined by means of two platinum resistance thermometers (S1059PJ5X6, Minco Products, Inc., Minneapolis MN, USA). A very accurate monitoring and measurement of temperature is achieved using an AC comparator resistance bridge (F700, Automatic Systems Laboratories, Redhill, England) connected to the Pt thermometers and a reference resistance. The pressure inside the cell is recorded by two pressure transducers which cover different pressure ranges: a Paroscientific 2500A-101 for pressures from (0 to 3) MPa and a Paroscientific 43KR-HHT-101 (Paroscientific Inc., Redmond WA, USA) for pressures up to 20 MPa.

### 2.3. Density measurement procedure

A detailed description of the measurement procedure in SSMSD is presented by Mondéjar et al. [14] and Lozano-Martín et al. [15] for our equipment; and by McLinden [18] and Richter et al. [19] for general aspects. Basically, the density of the fluid can be calculated from Eq. (1):

$$\rho_{\text{fluid}} = \frac{m_{\text{s0}} - m_{\text{sf}}}{V_{\text{s}}(T,p)} \qquad (1)$$

where the difference between the result of weighing the sinker in vacuum $m_{\text{s0}}$ and the result of weighing the sinker in the pressurized fluid $m_{\text{sf}}$ is related to the buoyancy force exerted on the sinker which is determined using a high-precision microbalance. $V_{\text{s}}$ $(T, p)$ is the volume of the sinker immersed in the fluid, whose dependence on temperature and pressure is accurately known [14].

The measurement procedure involves obtaining the buoyancy force by subtracting the reading of the balance in two different positions of the magnetic coupling, namely the zero position (ZP) and the measuring position (MP). In the ZP the magnetic coupling attracts the permanent magnet, but the sinker is not lifted and continues to rest on the bottom of the cell. In the MP, the magnetic coupling attracts the permanent magnet a little more strongly, so the sinker is lifted. Simultaneously, the load compensation system places the tantalum mass on the upper pan of the balance when the magnetic coupling is in the ZP, and it places the titanium mass when the magnetic coupling is in the MP. The differential nature of the measuring method cancels the weights of the permanent magnet, the electromagnet, the hook and sinker support, and their corresponding buoyancy forces. The air buoyancy on the titanium and tantalum compensation loads is approximately equal, as both loads have the same volume, and their difference is also cancelled. To achieve maximum accuracy, the calibration of the balance and the correction for the force transmission error (FTE) [20] must be considered. The FTE arises from the different perturbations of the magnetic materials, the external magnetic fields, and the magnetic behavior of the fluid being measured to the force transmitted through the magnetic coupling system. So, the final expression to obtain the experimental density is given by the more specific equation (2):

$$\rho_{\text{fluid}} = \frac{\Phi_0 m_{\text{s}} + (m_{\text{Ti}} - m_{\text{Ta}}) - (W_{\text{ZP}} - W_{\text{MP}})/\alpha}{V_{\text{s}}(T,p)} \frac{1}{\Phi_0} + \frac{\varepsilon_{\rho}}{\Phi_0} \frac{\chi_{\text{s}}}{\chi_{\text{s0}}} \left( \frac{\rho_{\text{s}}}{\rho_0} - \frac{\rho_{\text{fluid}}}{\rho_0} \right) \rho_{\text{fluid}} \qquad (2)$$

Equation (2) results from the applied method and thus should be explained in detail. The experimental density of the fluid is written as the sum of two terms. The first term stands for the density of the fluid if only the correction for the apparatus-specific effect of the FTE is considered. The second term of the sum corrects the experimental density considering the effect on the magnetic coupling due to the individual magnetic susceptibility of the fluid being measured, the fluid-specific effect of the FTE.

Regarding the first term of the sum in equation (2), $m_s$, $m_{Ti}$, and $m_{Ta}$ represent the masses of the silicon sinker and the titanium and tantalum compensation masses. ($W_{ZP}$ - $W_{MP}$) is the difference between the balance readings when the magnetic coupling is in the ZP and in the MP. $V_S$ ($T$, $p$) is the volume of the sinker at the experimental temperature and pressure. $\alpha$ and $\Phi_0$ are the balance calibration factor and the coefficient for the correction due to the apparatus-specific effect of the FTE (coupling factor).

The balance calibration factor $\alpha$ can easily be obtained by an independent calibration, free from the air buoyancy effect, using the two compensation masses. While the magnetic coupling is in the zero position (ZP), two measurements are recorded, one with the titanium mass on the balance pan ($W_{ZP,Ti}$), and the other with the tantalum mass ($W_{ZP,Ta}$). Subtracting one from the other gives the value of $\alpha$:

$$\alpha = \frac{W_{ZP,Ta} - W_{ZP,Ti}}{(m_{Ta} - m_{Ti})} \tag{3}$$

The balance calibration factor is measured directly before and after every single measurement point during the measurement campaign.

The term $\Phi_0$ accounts for the apparatus-specific effect of the FTE in the reading of the balance. It can be obtained by weighing the sinker in vacuum:

$$\Phi_0 = \frac{-(m_{Ti} - m_{Ta}) + (W_{ZP,vacuum} - W_{MP,vacuum})/\alpha}{m_s} \tag{4}$$

The apparatus-specific effect $\Phi_0$ is temperature-dependent and very sensitive to changes in the alignment of the electromagnet with respect to the permanent magnet, so it is experimentally determined at the end of every single isotherm to guarantee the highest accuracy of the measurement.

The second term on the right-hand side of equation (2) corrects the experimental density with the effect on the magnetic coupling due to the magnetic susceptibility of the fluid being measured (fluid-specific effect of the FTE). It should be pointed out that an approximate initial value of $\rho_{\text{fluid}}$ must be introduced to obtain this second term. The value of density which results from the first term of the right-hand side of that same equation is a good approximation for this purpose. The quantities $\chi_{s0} = 10^{-8}$ m$^3$·kg$^{-1}$ and $\rho_0 = 1000$ kg·m$^{-3}$ , which appear in equation (2), are reducing constants for the mass-based magnetic susceptibility of the fluid $\chi_s$, the density of the fluid $\rho_{\text{fluid}}$, and the density of the sinker $\rho_s$. The term $\varepsilon_\rho$, called the 'apparatus-specific constant of the fluid-specific effect', is specific for any particular SSMSD and was calculated for our equipment from density measurements on pure oxygen over a wide range of temperatures and pressures [15]. The obtained value for $\varepsilon_\rho$, as a function of temperature and density, is given by equation (5):

$$\varepsilon_\rho(T,\rho) = 8.822 \cdot 10^{-5} + 4.698 \cdot 10^{-8} \cdot (T/\mathrm{K} - 293.15) - 3.015 \cdot 10^{-8} \cdot \rho/(\mathrm{kg} \cdot \mathrm{m}^{-3}) \qquad (5)$$

The mass-based magnetic susceptibilities $\chi_s$ for the three ($CO_2$ + $O_2$) binary mixtures studied in this work were estimated using the additive law proposed by Bitter [21]:

$$\chi_s(T) = x_{\mathrm{CO_2}} \cdot \frac{M_{\mathrm{CO_2}}}{M_{\mathrm{mixture}}} \cdot \chi_{\mathrm{s,CO_2}} + x_{\mathrm{O_2}} \cdot \frac{M_{\mathrm{O_2}}}{M_{\mathrm{mixture}}} \cdot \chi_{\mathrm{s,O_2}}(T) \qquad (6)$$

where $x_{\mathrm{CO2}}$ and $x_{\mathrm{O2}}$ are the mole fractions of the components of the mixture and $M_{\mathrm{CO2}}$, $M_{\mathrm{O2}}$, and $M_{\mathrm{mixture}}$ are the molar mass of the components of the mixture and of the mixture itself. The mass-based magnetic susceptibility of carbon dioxide has a weak value of $\chi_{\mathrm{s,CO2}} = -0.6\cdot10^{-8}$ m$^3$·kg$^{-1}$ [22], with no temperature dependence, as is characteristic for diamagnetic fluids. The molar magnetic susceptibility of molecular oxygen at the reference state, $T$ = 293.15 K, $p$ = 0 MPa, and zero frequency, is taken as $\chi_{\mathrm{M00}} = (42.92 \pm 0.06)\cdot10^{-9}$ m$^3$·mol$^{-1}$, a value measured by May et al. [23] and in close accordance with the ab initio calculation of Minaev [24]. With $M_{\mathrm{O2}}$ = 31.9988 g·mol$^{-1}$ taken as the molar mass of oxygen, this gives a mass-based magnetic susceptibility of oxygen at the reference state of $\chi_{00} = 1.341\cdot10^{-6}$ m$^3$·kg$^{-1}$. Additionally, the dependence of the magnetic susceptibility of strong paramagnetic fluids, such as oxygen, with temperature should not be neglected [23]. The Curie law, which states that the magnetic susceptibility is proportional to $1/T$, has been used to account for this temperature dependence:

$$\chi_{\mathrm{s,O_2}}(T) = \chi_{00}\frac{293.15\ \mathrm{K}}{T} \tag{7}$$

Even though the magnetic susceptibility of pure oxygen may also have a weak dependence with density [23], resulting in a decrease in the magnetic susceptibility with increasing density; this effect is not to be considered here, as it is already taken into account with the density dependence of $\varepsilon_\rho$ reflected in equation (5) [15].

### 2.4. Experimental procedure

Experimental density data for the three ($CO_2$ + $O_2$) binary mixtures ($x(O_2)$ = 0.05, 0.10 and 0.20) were obtained at temperatures of (250, 260, 275, 293.15, 300, 325, 350, and 375) K and pressures up to 13 MPa. In the course of a measurement campaign for each isotherm, the pressure was reduced in 1 MPa steps, starting from the highest measured pressure down to 1 MPa. Figure 1 illustrates the recorded data as coordinates in a *p*, *T*-diagram together with the saturation curve for the mixture calculated with the EOS-CG [6]. The *p*, *T*-range of applicability of the EOS-CG and the main area of interest for CCS applications are also represented in the three plots of Figure 1.

Thirty repeated measurements are recorded for each single ($p$, $\rho$, $T$) point and the last ten values are used to obtain the mean value. The balance calibration factor $\alpha$ is obtained right before and after every single point, and the apparatus-specific effect $\Phi_0$ is determined at the end of every single isotherm.

Sorption effects inside the measuring cell may be responsible for errors of up to 0.1 % in density [19]. In order to minimize this effect, the measuring cell is evacuated and flushed several times with fresh mixture before each isotherm is measured, as recommended by Richter and Kleinrahm [19]. The residence time of the mixture in the cell is never longer than 40 hours. Specific sorption tests for this particular mixture were performed in the same way as they were done in previous investigations [11], [25]–[35]. Continuous density measurements on the same state point were recorded over 48 hours. These results showed that the difference observed in the trend of the relative deviation in density from the GERG-2008 EoS between the first and the last measurement is one order of magnitude lower than the density uncertainty. A measurement with fresh mixture executed immediately afterwards, for the same temperature and pressure, also repeated the density value with a deviation of one order of magnitude lower than the density uncertainty. Therefore, residual errors due to sorption effects are not appreciable with the experimental technique, and it should be considered that they are already included in the measurement uncertainty of the density and in the uncertainty of the composition.

The performance of the experimental technique and the measurement method was checked before and after the measurements of the three ($CO_2$ + $O_2$) mixtures, using nitrogen as the reference fluid over the entire operational range of the apparatus. These results were compared to the densities calculated from the reference equation of state for nitrogen by Span et al. [36]. The relative deviations of the 113 experimental densities of nitrogen, measured over seven different isotherms from 240 K to 350 K and for pressures up to 20 MPa, agreed with those calculated from the reference equation of state as the relative deviations remain within a ±0.02 % band, with an absolute average deviation (AAD) of 0.014 %.

## 2.5. Uncertainty of the measurements

A detailed analysis of the uncertainties of the measurements involved in this experimental procedure was reported in previous works [14][15]. The expanded uncertainty in temperature ($k$ = 2) is less than 4 mK. The pressure uncertainty depends on the range and is given by Eq. (8) and Eq. (9) for the (3 to 20) MPa and (0 to 3) MPa transducers, respectively. The expanded uncertainty ($k$ = 2) in pressure is in both cases less than 0.005 MPa.

$$U(p)/\text{MPa} = 75\cdot 10^{-6}\cdot p/\text{MPa} + 3.5\cdot 10^{-3} \tag{8}$$

$$U(p)/\text{MPa} = 60\cdot 10^{-6}\cdot p/\text{MPa} + 1.7\cdot 10^{-3} \tag{9}$$

The uncertainty of density data for the three ($CO_2$ + $O_2$) binary mixtures, corrected by both the apparatus-specific and the fluid-specific FTE effects, $U(\rho_{\text{fluid}})$, is evaluated directly by applying the law of propagation of uncertainty, according to the Guide to the Expression of Uncertainty in Measurement (GUM) [37], to equation (2). The estimation of the uncertainty was thoroughly explained in a previous paper [15] and can be given by the expression:

$$U(\rho)/\mathrm{kg{\cdot}m^{-3}} = 2.5{\cdot}10^{4}{\cdot}\chi_s/\mathrm{m^3{\cdot}kg^{-1}} + 1.1{\cdot}10^{-4}{\cdot}\rho/\mathrm{kg{\cdot}m^{-3}} + 2.3{\cdot}10^{-2} \quad (10)$$

To calculate the overall expanded uncertainty in density $U_T(\rho)$ ($k$ = 2), the uncertainties of density, temperature, pressure, and composition of the mixture must be included, as expressed in Eq. (11):

$$U_\mathrm{T}(\rho) = 2 \cdot \left[ u(\rho)^2 + \left( \left(\frac{\partial \rho}{\partial p}\right)_{T,x} \cdot u(p) \right)^2 + \left( \left(\frac{\partial \rho}{\partial T}\right)_{p,x} \cdot u(T) \right)^2 + \sum_i \left( \left(\frac{\partial \rho}{\partial x_i}\right)_{T,p,x_j \neq x_i} \cdot u(x_i) \right)^2 \right]^{0.5} \quad (11)$$

where $p$ is the pressure, $T$ is the temperature, and $x_i$ is the amount-of-substance (mole) fraction of each component in the mixture. Partial derivatives were calculated from the GERG-2008 EoS using the REFPROP software [38].

Table 4 displays a summary of the uncertainty contributions of density, temperature, pressure, and composition to the overall uncertainty in density for the three studied ($CO_2$ + $O_2$) binary mixtures.

## 3. Experimental results

Tables 5, 6, and 7 show the 162 experimental ($p$, $\rho$, $T$) data measured for the three ($CO_2$ + $O_2$) binary mixtures. The temperature, pressure, and density of each measured point were calculated as the arithmetic mean of the last ten consecutive measurements of a series of thirty. Tables 5, 6, and 7 also show the expanded uncertainty in density $U(\rho_{\mathrm{exp}})$ ($k$ = 2), calculated by Eq. (10) and expressed in absolute density units and as a percentage of the measured density.

The experimental data were compared to the corresponding densities calculated from the GERG-2008 and the EOS-CG equations of state, using the REFPROP [38] and TREND 4.0 [39] software (with the original, short EoS used in the GERG model for pure fluids, instead of the more precise and more complex reference EoS for pure fluids used in REFPROP by default). The relative deviations of the experimental densities from these EoS are included in Tables 5, 6, and 7 and are shown in Figures 2, 3, and 4.

It is worth mentioning that the densities of the experimental points recorded in this work range from $\rho$ = 9.843 kg·m$^{-3}$ ($T$ = 260 K, $p$ = 0.5 MPa, $x(O_2)$ = 0.20) to $\rho$ = 478.56 kg·m$^{-3}$ ($T$ = 300 K, $p$ = 12.3 MPa, $x(O_2)$ = 0.20).

It is also worth pointing out the relevance of the correction due to the fluid-specific effect of the FTE, which can be applied from the estimation of the apparatus-specific constant $\varepsilon_\rho$ of the fluid-specific effect in a previous work [15]. This correction is represented by the second term of the right-hand side of Eq. (2), and should be considered for any mixture with a high content of any paramagnetic fluid, such as oxygen, and increases with the density of the measured fluid. The correction due to the fluid-specific effect in the mixtures measured in this work can be as high as 1.313 kg·m$^{-3}$ in absolute value (0.28 % relative value) at the high density of $\rho$ = 494.505 kg·m$^{-3}$, or 0.47 % in relative values (0.049 kg·m$^{-3}$ absolute value) at the lowest density of $\rho$ = 10.322 kg·m$^{-3}$. In both cases, this applies to the mixture with the higher oxygen content (0.80 $CO_2$ + 0.20 $O_2$) at $T$ = 300 K and $T$ = 250 K, respectively.

## 4. Discussion of the results

Figure 2 shows the relative deviations of the experimentally determined density data of the (0.95 $CO_2$ + 0.05 $O_2$) mixture from the corresponding density data calculated by the GERG-2008 (a) and the EOS-CG (b) models. In the same way, Figures 3 and 4 show the deviations for the (0.90 $CO_2$ + 0.10 $O_2$) mixture and the (0.80 $CO_2$ + 0.20 $O_2$) mixture, respectively.

Both equations of state claim an uncertainty in density of 1.0 % for mixtures of $CO_2$ and $O_2$, over the temperature range from (250 to 450) K and at pressures up to 35 MPa. The estimated uncertainty of the experimental density data ranges from 0.017 %, for $T$ = 300 K, $p$ = 12.3 MPa ($\rho$ = 478.56 kg·m$^{-3}$), to 0.211 % for $T$ = 375 K, $p$ = 1.0 MPa ($\rho$ = 13.550 kg·m$^{-3}$), in both cases for the (0.80 $CO_2$ + 0.20 $O_2$) mixture. A slightly bigger relative uncertainty of 0.3 % can be found in two single points, for the (0.80 $CO_2$ + 0.20 $O_2$) mixture, at $T$ = 250 K and 260 K. This is because these are the only two points that have been measured at the lowest pressure of $p$ = 0.5 MPa and show the lowest densities within this study ($\rho$ = 9.843 kg·m$^{-3}$ and $\rho$ = 10.322 kg·m$^{-3}$). Thus, for most of the measurements, the experimental expanded ($k$ = 2) uncertainty is two orders of magnitude lower than the stated uncertainty of both equation-of-state models; and, even at the lowest experimental density, the estimated experimental uncertainty is still three times lower.

The relative deviations of the experimental density data from those calculated by GERG-2008 for the (0.95 $CO_2$ + 0.05 $O_2$) mixture (Figure 2 (a)) remain within the claimed uncertainty of the equation of state for all the 45 experimental points except for three. These three points correspond to the highest measured pressure of each of the three lowest temperatures measured ($T$ = 275 K and $p$ = 4.0 MPa; $T$ = 293.15 K and $p$ = 5.0 MPa; and $T$ = 300 K and $p$ = 6.0 MPa). The relative deviations for these three points are between 1.2 % and 1.5 %. The deviations from this equation-of-state model increase as the oxygen content in the mixture increases. For the (0.90 $CO_2$ + 0.10 $O_2$) mixture, 10 of the 47 experimental points already deviate by more than the claimed uncertainty of the GERG-2008 EoS (Figure 3 (a)). These points belong to the highest pressures measured at temperatures of 275 K, 293.15 K, 300 K and 325 K. The relative deviations can become as high as 4.4 %. For the (0.80 $CO_2$ + 0.20 $O_2$) mixture, 31 of the 70 measured points deviate by more than the claimed uncertainty of the GERG-2008 EoS (Figure 4 (a)). This behavior depends on the temperature, as the points are located outside the margin for pressures over 2 MPa at $T$ = 250 and 260 K, over 3 MPa at $T$ = 275 K, over 4 MPa at $T$ = 293.15 and 300 K, over 6 MPa at $T$ = 325 K, over 9 MPa at $T$ = 350 K, and over 12 MPa at $T$ = 375 K. The relative deviations can amount to as much as 6.6 %. In this mixture, we observe a non-monotonous run with increasing pressures. At $T$ = 300 K, the relative deviation goes through a distinct maximum at approximately 10 MPa. Generally speaking, the GERG-2008 EoS can fit the experimental data within its claimed uncertainty only for the mixture with a lower oxygen content. The deviations visibly increase when the oxygen content increases, , this being particularly pronounced at lower temperatures and higher pressures. Furthermore, the deviations always have a positive value, i.e., the GERG-2008 underestimates the density of ($CO_2$ + $O_2$) mixtures, notably for mixtures with a high oxygen content, at high pressures and low temperatures.

The relative deviations of the experimental density data from those calculated by the other model, the EOS-CG (Figures 2 (b), 3 (b) and 4 (b)), for all the 162 experimental points, remain within the claimed uncertainty of the equation of state, except for three points, namely at ($x(O_2)$ = 0.10, $T$ = 293.15 K, $p$ = 6.0 MPa), ($x(O_2)$ = 0.20, $T$ = 300 K, $p$ = 11.0 MPa), and ($x(O_2)$ = 0.20, $T$ = 300 K, $p$ = 12.2 MPa), where the relative deviation increases up to 1.2 %, -2.0 % and -3.2 %, respectively.

Table 8 presents the statistical parameters of the relative deviation of the experimental data from the corresponding densities calculated by the GERG-2008 and the EOS-CG model. The absolute average deviation (AAD) of the experimental data from the densities calculated by the GERG-2008 EoS is of 0.29 % for the (0.95 $CO_2$ + 0.05 $O_2$) mixture, 0.69 % for the (0.90 $CO_2$ + 0.10 $O_2$) mixture, and 1.30 % for the (0.80 $CO_2$ + 0.20 $O_2$) mixture. The corresponding AAD of the experimental data from the densities calculated by the EOS-CG model are 0.08 %, 0.15 %, and 0.22 %. For the three mixtures investigated in this study, all the statistical parameters that have been evaluated from the data processing are smaller for the EOS-CG model than the corresponding parameters for the GERG-2008 model.

The availability of data for ($CO_2$ + $O_2$) mixtures in the literature is limited to oxygen contents below $x(O_2)$ = 0.15 [40][41][42][43]. The statistical parameters of the relative deviation of these experimental data from the results of the two equation-of-state models are also given in Table 8.

## 5. Conclusions

New ($p$, $\rho$, $T$) high-precision experimental data for three binary mixtures of carbon dioxide and oxygen, with nominal compositions of (0.95 $CO_2$ + 0.05 $O_2$), (0.90 $CO_2$ + 0.10 $O_2$), and (0.80 $CO_2$ + 0.20 $O_2$) at temperatures between (250 and 375) K and pressures up to 13 MPa, are reported. The experimental device used was a single-sinker densimeter with magnetic suspension coupling. The mixtures were prepared gravimetrically, which qualifies them as metrologically traceable reference mixtures.

The new experimental data were compared to the corresponding densities calculated by two equation-of-state models, the EOS-CG, which is specifically designed for combustion gases, and the GERG-2008, which is the established approach for working with natural gases. Both models claim an uncertainty margin of 1 % for the $p$, $T$-range investigated. The EOS-CG shows a good performance for the compositions investigated; the relative deviations in density were found to remain within the uncertainty margin of 1 % for almost all the data. The corresponding results from processing the GERG-2008, however, do not reach the same level as the results from the EOS-CG. The processed data show a significant deviation to positive values that becomes more pronounced towards lower temperatures, higher pressures, and higher oxygen contents. The deviation might be as high as about 7 % at 300 K and 10 MPa at $x(O_2)$ = 0.20. It seems that the underlying basis still needs further input and these results might direct attention to mixtures with oxygen. Since the investigation of mixtures with a significant amount of oxygen is also a challenge from the point of view of safety, appropriate measures are required.

## Acknowledgments

The authors wish to thank the Ministerio de Economía, Industria y Competitividad for their support through the project ENE2017-88474-R and the Junta de Castilla y León project VA280P18. The second author is grateful to the members of his affiliation group and program Erasmus + KA107-36589 ICM-UVa.

## Figures

**Figure 1.** *p*, *T*-phase diagram showing the experimental points measured (•) and the calculated phase envelope (solid line) using the EOS-CG [6] for: (a) (0.95 $CO_2$ + 0.05 $O_2$), (b) (0.90 $CO_2$ + 0.10 $O_2$), and (c) (0.80 $CO_2$ + 0.20 $O_2$) binary mixtures, respectively. The marked temperature and pressure ranges represent the range of the binary experimental data used for the development of the GERG-2008 [3] (blue dashed line), and the EOS-CG [6] (red dashed line), respectively, and the area of interest for the gas industry (black thin dashed line).

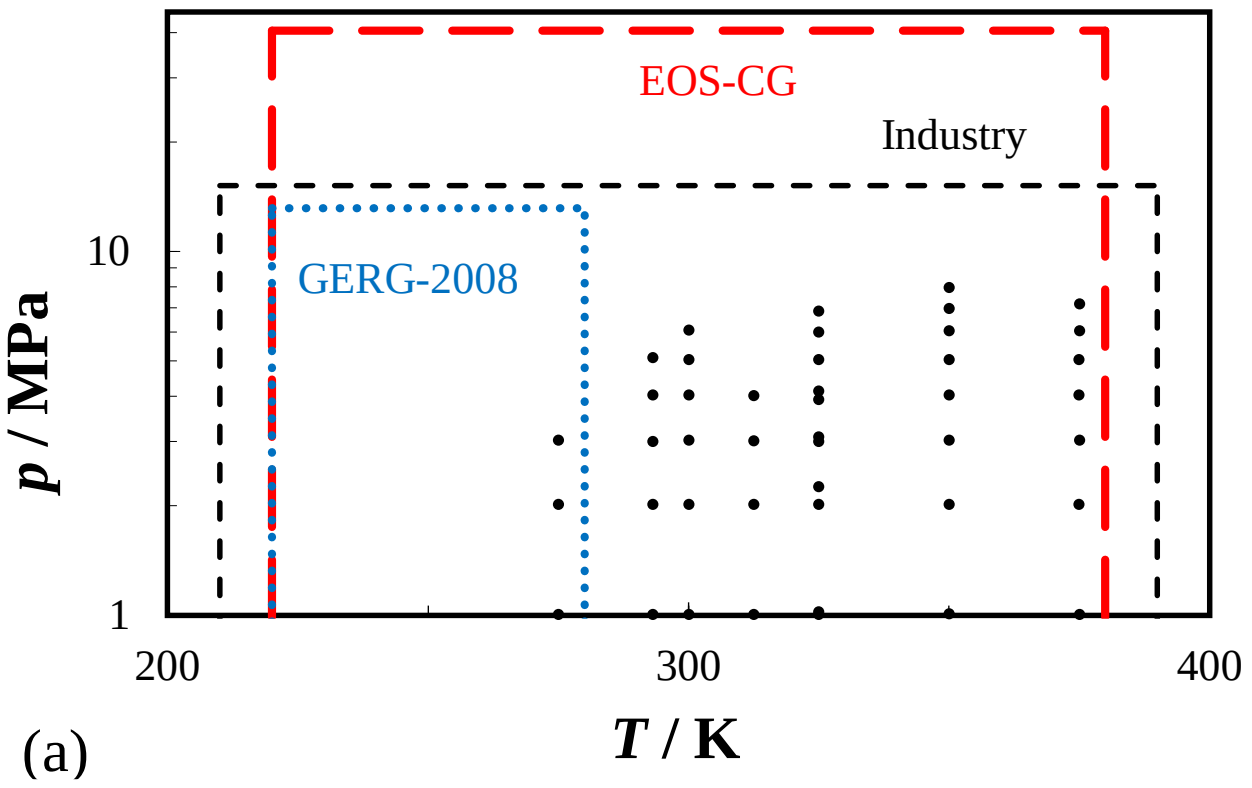

(a)

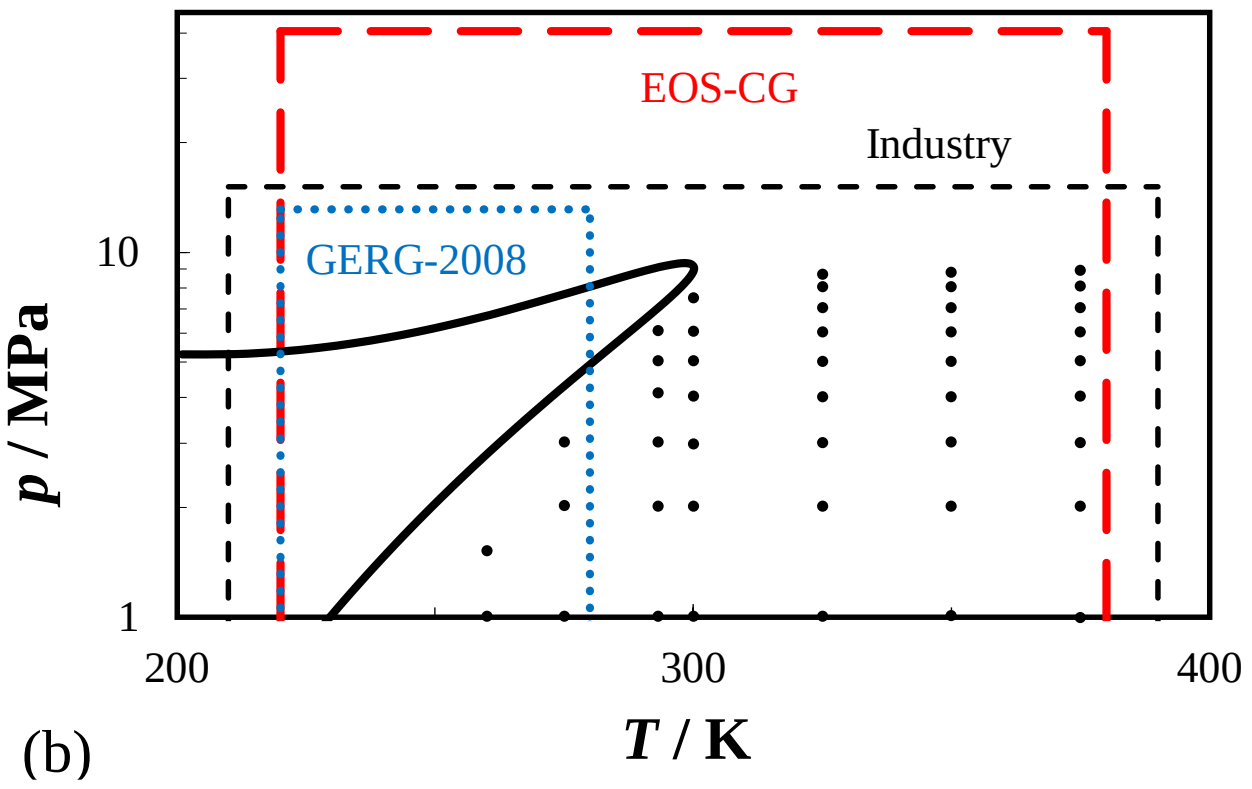

(b)

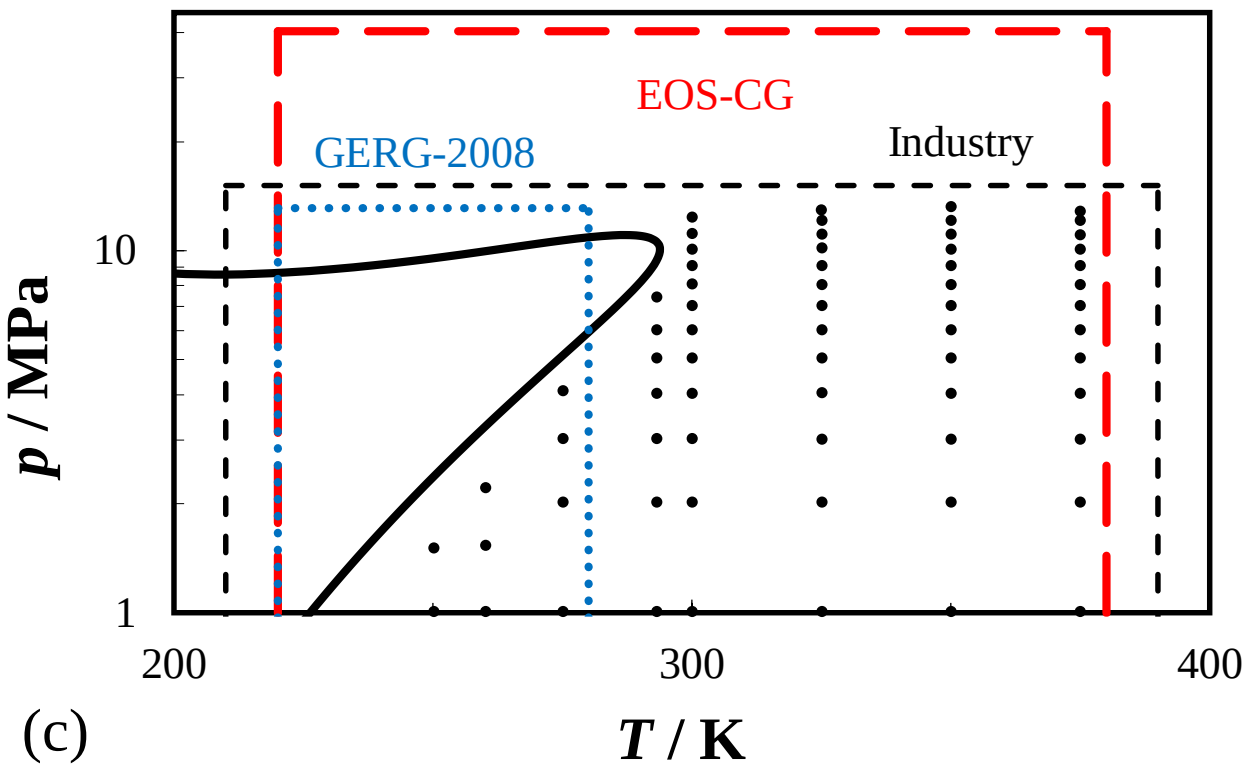

(c)

**Figure 2.** Relative deviations in density of experimental ($p$, $\rho_{exp}$, $T$) data of the binary (0.95 $CO_2$ + 0.05 $O_2$) mixture from density values calculated by the: (a) GERG-2008 [3], $\rho_{GERG}$, and (b) EOS-CG [6], $\rho_{CG}$, equations of state as a function of pressure for different temperatures: △ 275 K, □ 293.15 K, ○ 300 K, ◇ 312.5 K, + 325 K, ✳ 350 K, — 375 K. Dashed lines indicate the expanded ($k$ = 2) uncertainty of the corresponding EoS. Error bars on the 293.15 K data set indicate the expanded ($k$ = 2) uncertainty of the experimental density.

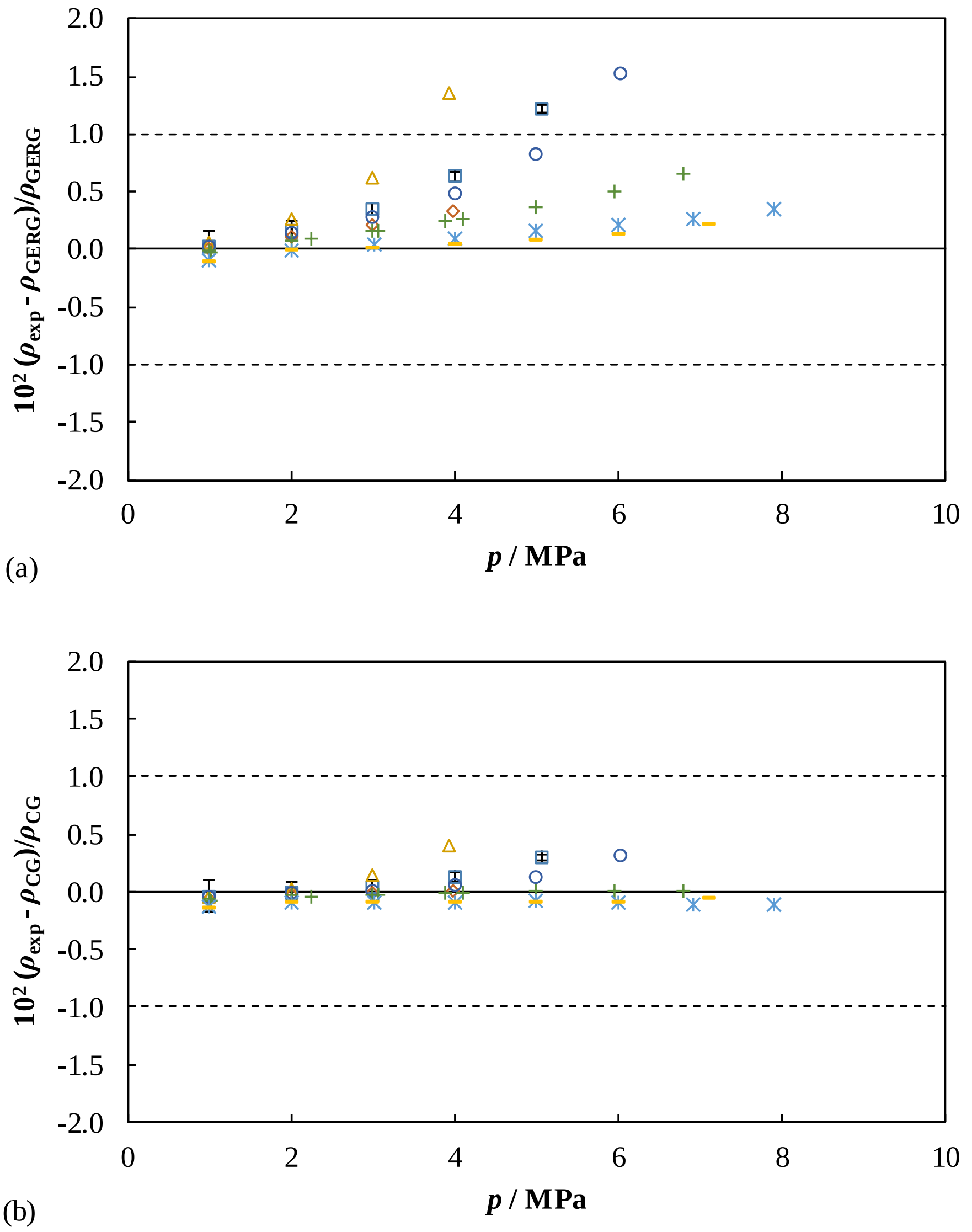

**Figure 3.** Relative deviations in density of experimental ($p$, $\rho_{exp}$, $T$) data of the binary (0.90 $CO_2$ + 0.10 $O_2$) mixture from density values calculated by the: (a) GERG-2008 [3], $\rho_{GERG}$, and (b) EOS-CG [6], $\rho_{CG}$, equations of state as a function of pressure for different temperatures: ◇ 260 K, × 275 K, □ 293.15 K, ○ 300 K, + 325 K, ✳ 350 K, — 375 K. Dashed lines indicate the expanded ($k$ = 2) uncertainty of the corresponding EoS. Error bars on the 293.15 K data set indicate the expanded ($k$ = 2) uncertainty of the experimental density.

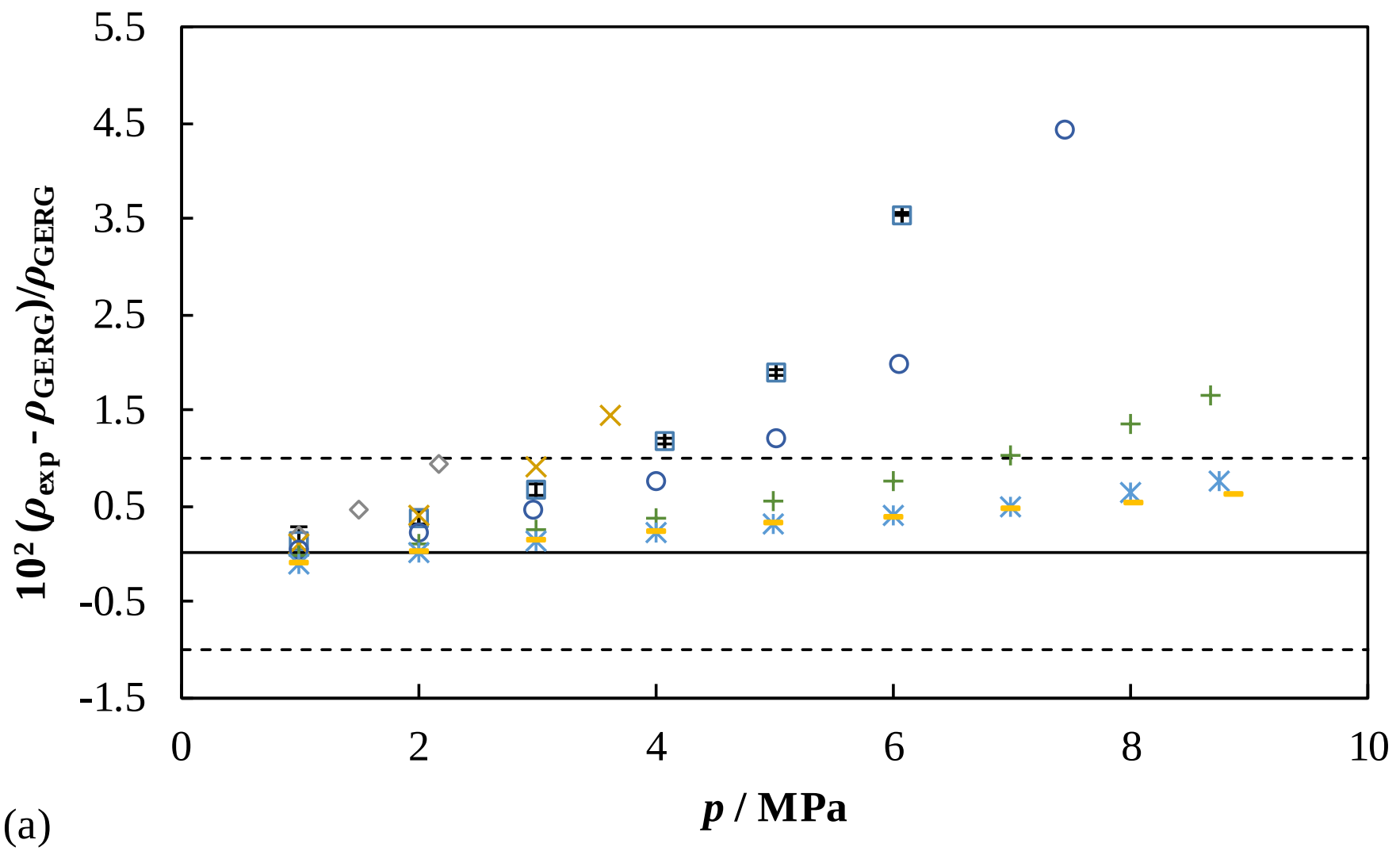

(a)

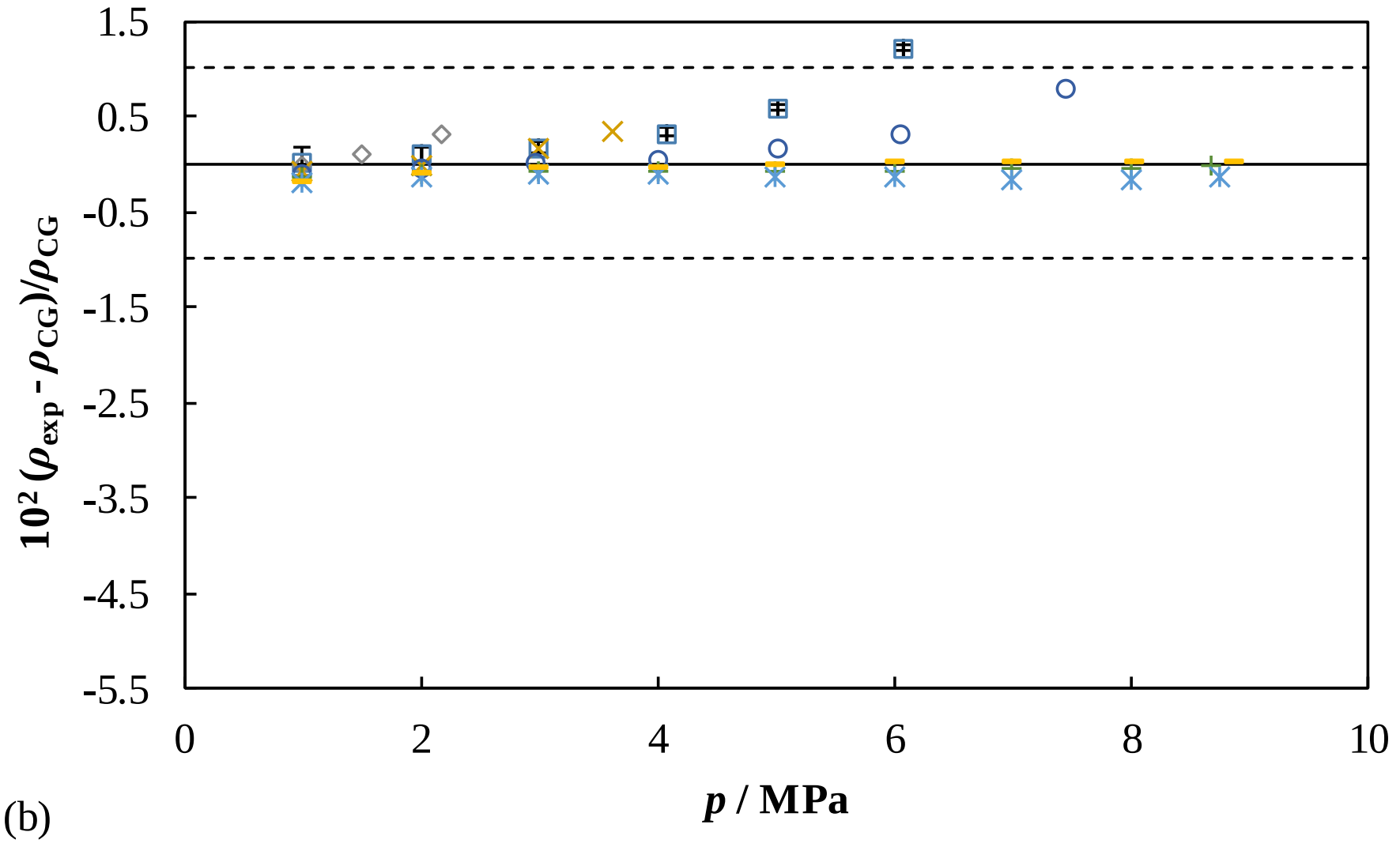

(b)

**Figure 4.** Relative deviations in density of experimental ($p$, $\rho_{exp}$, $T$) data of the binary (0.80 $CO_2$ + 0.20 $O_2$) mixture from density values calculated by the: (a) GERG-2008 [3], $\rho_{GERG}$, and (b) EOS-CG [6], $\rho_{CG}$, equations of state as a function of pressure for different temperatures: ◇ 250 K, △ 260 K, × 275 K, □ 293.15 K, ○ 300 K, + 325 K, ✳ 350 K, — 375 K. Dashed lines indicate the expanded ($k$ = 2) uncertainty of the corresponding EoS. Error bars on the 293.15 K data set indicate the expanded ($k$ = 2) uncertainty of the experimental density.

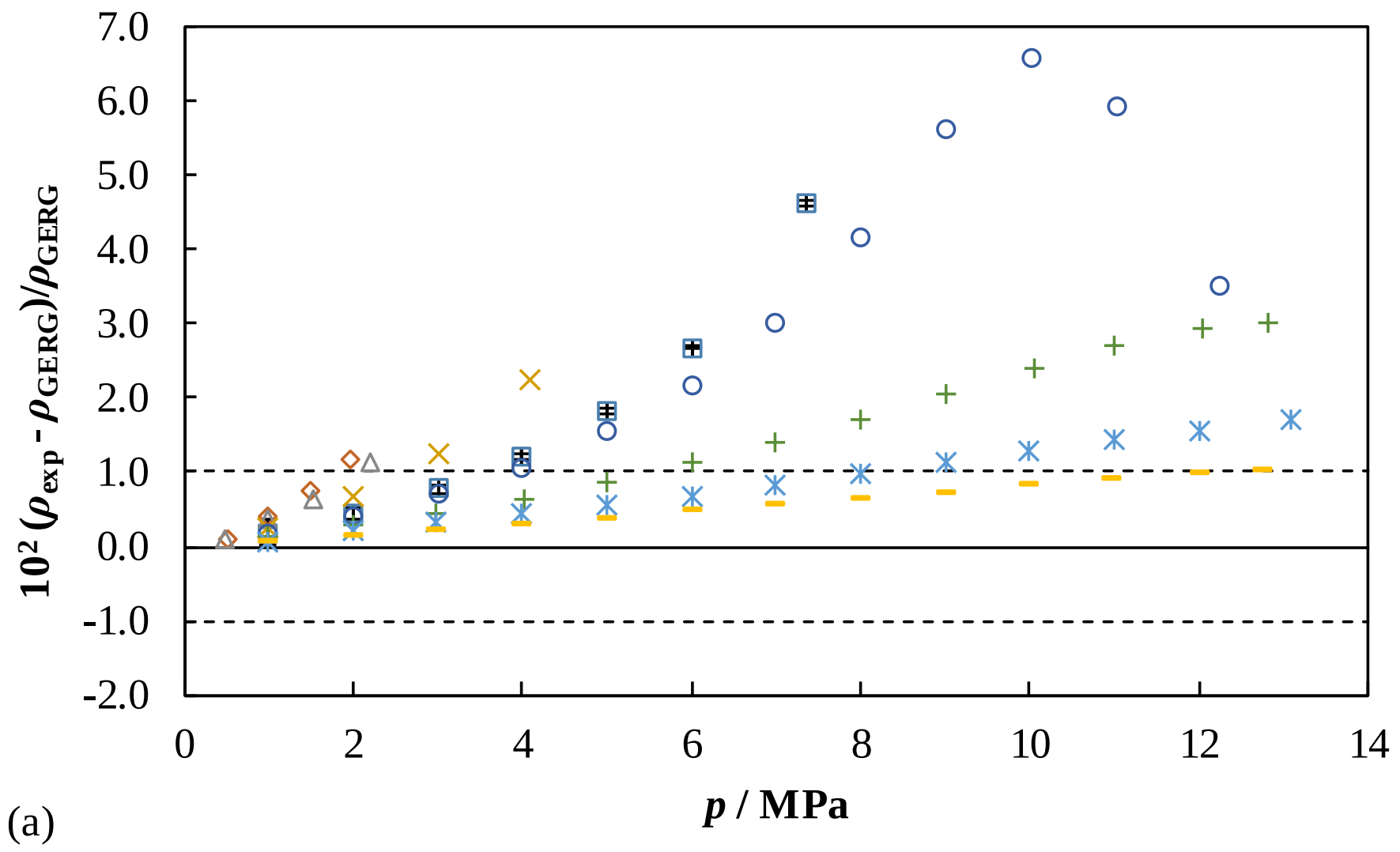

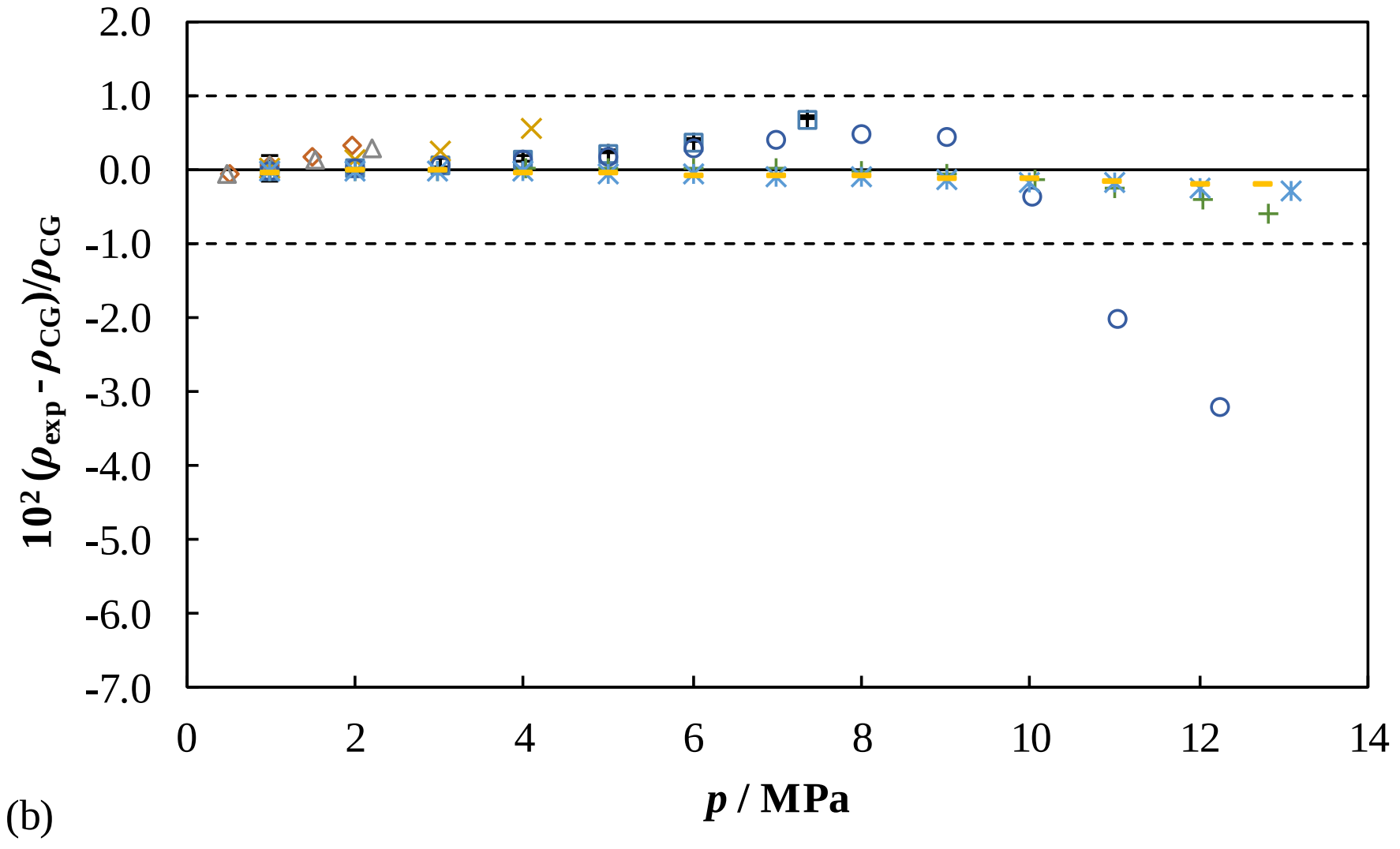

**Tables**

**Table 1.** Purity, supplier, molar mass and critical parameters of the constituent components of the studied ($CO_2$ + $O_2$) mixtures in this work.

| | Purity / vol-% | Supplier | $M$ / g·mol$^{-1}$ | Critical parameters[a] | |
|---|---|---|---|---|---|
| | | | | $T_c$ / K | $p_c$ / MPa |
| Carbon dioxide | 99.9995 | Air Liquide | 44.010 | 304.13 | 7.3773 |
| Oxygen | 99.9999 | Linde | 31.999 | 154.58 | 5.0430 |

[a] Critical parameters were obtained by using the default equation for each substance in REFPROP software [38].

**Table 2.** Composition with its expanded ($k$ = 2) uncertainty of the studied binary ($CO_2$ + $O_2$) mixtures in this work. Impurity compounds are marked in *italics*.

| Component | $(0.95\ CO_2 + 0.05\ O_2)^{(a)}$ | | $(0.90\ CO_2 + 0.10\ O_2)^{(b)}$ | | $(0.80\ CO_2 + 0.20\ O_2)^{(c)}$ | |
|---|---|---|---|---|---|---|
| | $10^2\ x_i$ / mol/mol | $10^2\ U(x_i)$ / mol/mol | $10^2\ x_i$ / mol/mol | $10^2\ U(x_i)$ / mol/mol | $10^2\ x_i$ / mol/mol | $10^2\ U(x_i)$ / mol/mol |
| Carbon dioxide | 94.967781 | 0.001312 | 90.014271 | 0.001190 | 80.009178 | 0.001121 |
| Oxygen | 5.032082 | 0.001800 | 9.985591 | 0.001632 | 19.990681 | 0.001538 |
| *Argon* | 0.000005 | 0.000006 | 0.000010 | 0.000012 | 0.000020 | 0.000023 |
| *Nitrogen* | 0.000098 | 0.000110 | 0.000096 | 0.000105 | 0.000090 | 0.000094 |
| *Carbon monoxide* | 0.000029 | 0.000033 | 0.000028 | 0.000031 | 0.000026 | 0.000028 |
| *Propane* | < 0.000001 | < 0.000001 | 0.000001 | 0.000001 | 0.000001 | 0.000002 |
| *Nitric oxide* | 0.000004 | 0.000005 | 0.000004 | 0.000005 | 0.000004 | 0.000004 |
| Normalized composition without impurities | | | | | | |
| Carbon dioxide | 94.967911 | 0.001312 | 90.014396 | 0.001190 | 80.009291 | 0.001121 |
| Oxygen | 5.032089 | 0.001800 | 9.985604 | 0.001632 | 19.990709 | 0.001538 |

(a) BAM cylinder no.: 1007-180626

(b) BAM cylinder no.: 1008-180626

(c) BAM cylinder no.: 9085-180116

**Table 3.** Results of the gas chromatographic (GC) analysis with its expanded ($k = 2$) uncertainty, and relative deviations between gravimetric preparation and GC analysis for the three ($CO_2$ + $O_2$) mixtures studied in this work. The results are followed by the gravimetric composition (non-normalized) of the validation mixtures employed.

| Component | Concentration | | Relative deviation between gravimetric composition and GC analysis |
|---|---|---|---|
| | $10^2\ x_i$ / mol/mol | $10^2\ U(x_i)$ / mol/mol | % |
| (0.95 $CO_2$ + 0.05 $O_2$) BAM cylinder no.: 1007-180626 | | | |
| Carbon dioxide | n. a. | n. a. | — |
| Oxygen | 5.0054 | 0.0292 | –0.530 |
| Validation mixture BAM cylinder no.: 8009-141027 (premixture of G 473) | | | |
| Carbon dioxide | 51.448818 | 0.000950 | |
| Nitrogen | 44.142855 | 0,001898 | |
| Oxygen | 4.408289 | 0.000340 | |
| *Argon* | 0.000009 | 0.000010 | |
| *Carbon monoxide* | 0.000018 | 0.000018 | |
| *Methane* | 0.000001 | 0.000001 | |
| *Hydrogen* | 0.000007 | 0.000006 | |
| *Nitric oxide* | 0.000002 | 0.000003 | |
| Validation mixture BAM cylinder no.: 1005-180528 (premixture of G 471) | | | |
| Carbon dioxide | 36.238235 | 0.000802 | |
| Nitrogen | 57.952277 | 0.001260 | |
| Oxygen | 5.809458 | 0.001102 | |
| *Argon* | 0.000006 | 0.000007 | |
| *Carbon monoxide* | 0.000014 | 0.000013 | |
| *Propane* | < 0.000001 | < 0.000001 | |
| *Hydrogen* | 0.000008 | 0.000009 | |
| *Nitric oxide* | 0.000002 | 0.000002 | |
| ($0.90\ CO_2 + 0.10\ O_2$) BAM cylinder no.: 1008-180626 | | | |
| Carbon dioxide | n. a. | n. a. | — |
| Oxygen | 9.9858 | 0.0036 | 0.002 |
| Validation mixture BAM cylinder no.: 1028-190729 (G 033) | | | |
| Carbon dioxide | 90.460031 | 0.001522 | |
| Oxygen | 9.539830 | 0.002090 | |
| *Argon* | 0.000010 | 0.000011 | |
| *Nitrogen* | 0.000095 | 0.000105 | |
| *Carbon monoxide* | 0.000028 | 0.000031 | |
| *Propane* | 0.000001 | 0.000001 | |
| *Nitric oxide* | 0.000005 | 0.000005 | |
| Validation mixture BAM cylinder no.: 1029-190729 (G 033) | | | |
| Carbon dioxide | 89.489623 | 0.001524 | |
| Oxygen | 10.510238 | 0.002094 | |
| *Argon* | 0.000011 | 0.000012 | |
| *Nitrogen* | 0.000095 | 0.000104 | |
| *Carbon monoxide* | 0.000028 | 0.000031 | |
| *Propane* | 0.000001 | 0.000001 | |
| *Nitric oxide* | 0.000004 | 0.000005 | |
| ($0.80\ CO_2 + 0.20\ O_2$) BAM cylinder no.: 9085-180116 | | | |
| Carbon dioxide | n. a. | n. a. | — |
| Oxygen | 19.9924 | 0.0048 | 0.009 |
| Validation mixture BAM cylinder no.: 9097-180205 (G 033) | | | |
| Carbon dioxide | 81.017404 | 0.001269 | |
| Oxygen | 18.982455 | 0.001742 | |
| *Argon* | 0.000019 | 0.000022 | |

| | | |
|---|---|---|
| *Nitrogen* | 0.000091 | 0.000095 |
| *Carbon monoxide* | 0.000026 | 0.000028 |
| *Propane* | 0.000001 | 0.000002 |
| *Nitric oxide* | 0.000004 | 0.000004 |
| Validation mixture BAM cylinder no.: 9099-180205 (G 033) | | |
| Carbon dioxide | 79.011446 | 0.001265 |
| Oxygen | 20.988412 | 0.001737 |
| *Argon* | 0.000021 | 0.000025 |
| *Nitrogen* | 0.000090 | 0.000093 |
| *Carbon monoxide* | 0.000026 | 0.000027 |
| *Propane* | 0.000002 | 0.000002 |
| *Nitric oxide* | 0.000003 | 0.000004 |

**Table 4.** Contributions to the expanded ($k$ = 2) overall uncertainty in density, $U_T(\rho_{exp})$, for the three ($CO_2$ + $O_2$) mixtures studied in this work.

| Source | Contribution ($k$ = 2) | Units | Estimation in density ($k$ = 2) | |
|---|---|---|---|---|
| | | | kg·m$^{-3}$ | % |
| | | ($0.95\ CO_2 + 0.05\ O_2$) | | |
| Temperature, $T$ | 0.004 | K | < 0.010 | < 0.0058 |
| Pressure, $p$ | < 0.004 | MPa | (0.052 – 0.20) | (0.068 – 0.39) |
| Composition, $x_i$ | < 0.0004 | mol·mol$^{-1}$ | < 0.083 | < 0.0076 |
| Density, $\rho$ | (0.026 – 0.043) | kg·m$^{-3}$ | (0.026 – 0.043) | (0.026 – 0.18) |
| Sum | | | (0.058 – 0.21) | (0.075 – 0.41) |
| | | ($0.90\ CO_2 + 0.10\ O_2$) | | |
| Temperature, $T$ | 0.004 | K | < 0.020 | < 0.0088 |
| Pressure, $p$ | < 0.004 | MPa | (0.051 – 0.30) | (0.057 – 0.98) |
| Composition, $x_i$ | < 0.0004 | mol·mol$^{-1}$ | < 0.026 | < 0.011 |
| Density, $\rho$ | (0.027 – 0.052) | kg·m$^{-3}$ | (0.027 – 0.052) | (0.022 – 0.19) |
| Sum | | | (0.058 – 0.30) | (0.064 – 1.0) |
| | | ($0.80\ CO_2 + 0.20\ O_2$) | | |
| Temperature, $T$ | 0.004 | K | < 0.049 | < 0.012 |
| Pressure, $p$ | < 0.005 | MPa | (0.050 – 0.34) | (0.042 – 0.74) |
| Composition, $x_i$ | < 0.0004 | mol·mol$^{-1}$ | < 0.037 | < 0.0082 |
| Density, $\rho$ | (0.029 – 0.083) | kg·m$^{-3}$ | (0.029 – 0.083) | (0.017 – 0.30) |
| Sum | | | (0.057 – 0.35) | (0.049 – 0.80) |

**Table 5.** Experimental ($p$, $\rho_{exp}$, $T$) measurements for the gaseous binary (0.95 $CO_2$ + 0.05 $O_2$) mixture, absolute and relative expanded ($k$ = 2) uncertainty in density, $U(\rho_{exp})$, and relative deviations from the density given by the GERG-2008 [3], $\rho_{GERG}$, and the EOS-CG [6], $\rho_{CG}$, equations of state. Normalized composition of the gaseous mixture is given in Table 2.

| $T$ / K[(a)] | $p$ / MPa[(a)] | $\rho_{exp}$ / kg·m$^{-3}$ | $U(\rho_{exp})$ / kg·m$^{-3}$ | $10^2\ U(\rho_{exp})/\rho_{exp}$ | $10^2\ (\rho_{exp} - \rho_{GERG})/\rho_{GERG}$ | $10^2\ (\rho_{exp} - \rho_{CG})/\rho_{CG}$ |
|---|---|---|---|---|---|---|
| 275 K isotherm | | | | | | |
| 275.001 | 3.940 | 109.732 | 0.037 | 0.034 | 1.35 | 0.41 |
| 275.005 | 3.003 | 72.969 | 0.033 | 0.045 | 0.62 | 0.16 |
| 275.001 | 1.999 | 43.849 | 0.029 | 0.067 | 0.26 | 0.04 |
| 275.002 | 0.998 | 20.228 | 0.027 | 0.132 | 0.05 | -0.03 |
| 293.15 K isotherm | | | | | | |
| 293.100 | 5.066 | 133.334 | 0.039 | 0.030 | 1.22 | 0.30 |
| 293.101 | 3.999 | 93.138 | 0.035 | 0.037 | 0.63 | 0.13 |
| 293.098 | 2.983 | 63.642 | 0.032 | 0.050 | 0.34 | 0.05 |
| 293.099 | 1.999 | 39.803 | 0.029 | 0.072 | 0.17 | 0.01 |
| 293.094 | 0.999 | 18.728 | 0.026 | 0.141 | 0.03 | -0.04 |
| 300 K isotherm | | | | | | |
| 299.946 | 6.024 | 165.383 | 0.043 | 0.026 | 1.52 | 0.33 |
| 299.946 | 5.001 | 121.053 | 0.038 | 0.031 | 0.82 | 0.14 |
| 299.946 | 4.003 | 88.437 | 0.034 | 0.039 | 0.48 | 0.06 |
| 299.945 | 3.000 | 61.548 | 0.031 | 0.051 | 0.28 | 0.02 |
| 299.949 | 1.999 | 38.512 | 0.029 | 0.074 | 0.14 | < 0.01 |
| 299.949 | 0.999 | 18.226 | 0.026 | 0.145 | 0.02 | -0.04 |
| 312.5 K isotherm | | | | | | |
| 312.472 | 3.988 | 80.993 | 0.033 | 0.041 | 0.34 | 0.02 |
| 312.476 | 2.987 | 57.253 | 0.031 | 0.054 | 0.21 | < 0.01 |
| 312.475 | 2.001 | 36.462 | 0.028 | 0.078 | 0.11 | -0.02 |
| 312.474 | 0.999 | 17.385 | 0.026 | 0.151 | 0.01 | -0.05 |
| | | | 325 K isotherm | | | |
| 324.954 | 6.804 | 151.790 | 0.041 | 0.027 | 0.65 | 0.01 |
| 324.954 | 5.950 | 125.498 | 0.038 | 0.031 | 0.50 | 0.02 |
| 324.954 | 4.999 | 99.738 | 0.036 | 0.036 | 0.37 | 0.01 |
| 324.953 | 4.102 | 78.087 | 0.033 | 0.042 | 0.27 | < 0.01 |
| 324.953 | 3.882 | 73.095 | 0.033 | 0.044 | 0.24 | -0.01 |
| 324.953 | 3.066 | 55.564 | 0.031 | 0.055 | 0.16 | -0.02 |
| 324.953 | 2.986 | 53.926 | 0.030 | 0.056 | 0.16 | -0.01 |
| 324.952 | 2.243 | 39.219 | 0.029 | 0.073 | 0.09 | -0.03 |
| 324.955 | 2.000 | 34.625 | 0.028 | 0.081 | 0.08 | -0.02 |
| 324.952 | 1.014 | 16.891 | 0.026 | 0.155 | -0.03 | -0.07 |
| 324.953 | 0.997 | 16.594 | 0.026 | 0.157 | -0.01 | -0.06 |
| | | | 350 K isotherm | | | |
| 349.938 | 7.897 | 153.813 | 0.042 | 0.027 | 0.35 | -0.11 |
| 349.939 | 6.910 | 129.216 | 0.039 | 0.030 | 0.26 | -0.11 |
| 349.939 | 6.001 | 108.351 | 0.036 | 0.034 | 0.22 | -0.08 |
| 349.937 | 4.999 | 87.007 | 0.034 | 0.039 | 0.16 | -0.08 |
| 349.938 | 3.999 | 67.238 | 0.032 | 0.047 | 0.10 | -0.08 |
| 349.937 | 3.005 | 48.906 | 0.030 | 0.061 | 0.04 | -0.09 |
| 349.937 | 1.999 | 31.537 | 0.028 | 0.088 | -0.01 | -0.09 |
| 349.938 | 0.999 | 15.308 | 0.026 | 0.169 | -0.09 | -0.13 |
| | | | 375 K isotherm | | | |
| 374.925 | 7.119 | 117.407 | 0.037 | 0.032 | 0.21 | -0.05 |
| 374.922 | 5.997 | 96.019 | 0.035 | 0.036 | 0.13 | -0.08 |
| 374.921 | 4.998 | 78.021 | 0.033 | 0.042 | 0.07 | -0.10 |
| 374.921 | 3.999 | 60.906 | 0.031 | 0.051 | 0.04 | -0.09 |
| 374.923 | 2.999 | 44.608 | 0.029 | 0.065 | 0.01 | -0.09 |
| 374.920 | 1.999 | 29.064 | 0.027 | 0.094 | -0.02 | -0.08 |
| 374.922 | 0.999 | 14.188 | 0.026 | 0.181 | -0.11 | -0.14 |

[a] Expanded uncertainties ($k$ = 2): $U(p > 3)/\mathrm{MPa} = 75 \cdot 10^{-6} \cdot \frac{p}{\mathrm{MPa}} + 3.5 \cdot 10^{-3}$; $U(p < 3)/\mathrm{MPa} = 60 \cdot 10^{-6} \cdot \frac{p}{\mathrm{MPa}} + 1.7 \cdot 10^{-3}$; $U(T)$ = 4 mK; $\frac{U(\rho)}{\mathrm{kg{\cdot}m^{-3}}} = 2.5 \cdot 10^{4} \frac{\chi_{\mathrm{s}}}{\mathrm{m^{3}{\cdot}kg^{-1}}} + 1.1 \cdot 10^{-4} \cdot \frac{\rho}{\mathrm{kg{\cdot}m^{-3}}} + 2.3 \cdot 10^{-2}$.

**Table 6.** Experimental ($p$, $\rho_{exp}$, $T$) measurements for the gaseous binary (0.90 $CO_2$ + 0.10 $O_2$) mixture, absolute and relative expanded ($k$ = 2) uncertainty in density, $U(\rho_{exp})$, and relative deviations from the density given by the GERG-2008 [3], $\rho_{GERG}$, and the EOS-CG [6], $\rho_{CG}$, equations of state. Normalized composition of the gaseous mixture is given in Table 2.

| $T$ / K[(a)] | $p$ / MPa[(a)] | $\rho_{exp}$ / kg·m$^{-3}$ | $U(\rho_{exp})$ / kg·m$^{-3}$ | $10^2$ $U(\rho_{exp})/\rho_{exp}$ | $10^2$ $(\rho_{exp} - \rho_{GERG})/\rho_{GERG}$ | $10^2$ $(\rho_{exp} - \rho_{CG})/\rho_{CG}$ |
|---|---|---|---|---|---|---|
| | | | 260 K isotherm | | | |
| 260.020 | 2.184 | 52.302 | 0.032 | 0.061 | 0.93 | 0.33 |
| 260.019 | 1.508 | 33.644 | 0.030 | 0.088 | 0.45 | 0.12 |
| 260.017 | 0.998 | 21.279 | 0.028 | 0.133 | 0.18 | -0.01 |
| | | | 275 K isotherm | | | |
| 274.989 | 3.627 | 90.941 | 0.036 | 0.040 | 1.43 | 0.34 |
| 274.984 | 2.999 | 70.068 | 0.034 | 0.048 | 0.90 | 0.17 |
| 274.980 | 2.008 | 42.944 | 0.031 | 0.071 | 0.39 | 0.01 |
| 274.976 | 0.998 | 19.835 | 0.028 | 0.141 | 0.09 | -0.06 |
| | | | 293.15 K isotherm | | | |
| 293.070 | 6.067 | 171.349 | 0.045 | 0.026 | 3.55 | 1.21 |
| 293.066 | 5.010 | 123.409 | 0.040 | 0.032 | 1.90 | 0.60 |
| 293.063 | 4.087 | 92.196 | 0.036 | 0.039 | 1.17 | 0.34 |
| 293.065 | 3.004 | 62.381 | 0.033 | 0.052 | 0.67 | 0.18 |
| 293.069 | 1.999 | 38.950 | 0.030 | 0.077 | 0.38 | 0.11 |
| 293.069 | 0.999 | 18.430 | 0.028 | 0.150 | 0.14 | 0.02 |
| | | | 300 K isotherm | | | |
| 299.924 | 7.455 | 231.402 | 0.052 | 0.022 | 4.43 | 0.81 |
| 299.926 | 6.044 | 153.702 | 0.043 | 0.028 | 1.99 | 0.33 |
| 299.925 | 5.017 | 115.524 | 0.039 | 0.033 | 1.22 | 0.17 |
| 299.924 | 4.001 | 85.165 | 0.035 | 0.041 | 0.75 | 0.07 |
| 299.924 | 2.968 | 59.083 | 0.032 | 0.055 | 0.46 | 0.03 |
| 299.923 | 1.999 | 37.687 | 0.030 | 0.079 | 0.22 | -0.03 |
| 299.924 | 0.999 | 17.908 | 0.028 | 0.154 | 0.03 | -0.08 |
| | | | 325 K isotherm | | | |
| 324.938 | 8.665 | 206.873 | 0.049 | 0.024 | 1.66 | < 0.01 |
| 324.938 | 7.994 | 182.298 | 0.046 | 0.025 | 1.36 | -0.03 |
| 324.938 | 7.002 | 150.100 | 0.042 | 0.028 | 1.03 | -0.03 |
| 324.938 | 6.000 | 121.618 | 0.039 | 0.032 | 0.75 | -0.05 |
| 324.937 | 5.000 | 96.328 | 0.036 | 0.038 | 0.54 | -0.06 |
| 324.937 | 3.999 | 73.596 | 0.034 | 0.046 | 0.37 | -0.06 |
| 324.937 | 2.986 | 52.660 | 0.031 | 0.059 | 0.24 | -0.06 |
| 324.936 | 2.000 | 33.932 | 0.029 | 0.086 | 0.10 | -0.08 |
| 324.937 | 0.999 | 16.347 | 0.027 | 0.166 | -0.03 | -0.11 |
| | | | 350 K isotherm | | | |
| 349.926 | 8.743 | 168.502 | 0.044 | 0.026 | 0.75 | -0.13 |
| 349.925 | 7.991 | 149.760 | 0.042 | 0.028 | 0.63 | -0.14 |
| 349.924 | 6.999 | 126.611 | 0.040 | 0.031 | 0.48 | -0.15 |
| 349.923 | 6.006 | 105.053 | 0.037 | 0.035 | 0.39 | -0.13 |
| 349.924 | 5.000 | 84.653 | 0.035 | 0.041 | 0.29 | -0.11 |
| 349.925 | 3.999 | 65.634 | 0.033 | 0.050 | 0.21 | -0.09 |
| 349.925 | 2.999 | 47.789 | 0.031 | 0.064 | 0.12 | -0.10 |
| 349.925 | 2.001 | 30.997 | 0.029 | 0.093 | 0.02 | -0.11 |
| 349.924 | 1.001 | 15.092 | 0.027 | 0.178 | -0.10 | -0.17 |
| | | | 375 K isotherm | | | |
| 374.915 | 8.878 | 148.273 | 0.042 | 0.028 | 0.60 | 0.01 |
| 374.915 | 8.034 | 131.519 | 0.040 | 0.030 | 0.53 | 0.02 |
| 374.913 | 6.993 | 111.704 | 0.038 | 0.034 | 0.45 | 0.01 |
| 374.913 | 6.009 | 93.842 | 0.036 | 0.038 | 0.38 | 0.02 |
| 374.912 | 5.001 | 76.325 | 0.034 | 0.044 | 0.29 | < 0.01 |
| 374.912 | 4.001 | 59.713 | 0.032 | 0.053 | 0.21 | -0.02 |
| 374.912 | 2.986 | 43.583 | 0.030 | 0.069 | 0.12 | -0.05 |
| 374.912 | 1.999 | 28.563 | 0.028 | 0.099 | 0.02 | -0.08 |
| 374.914 | 0.992 | 13.874 | 0.027 | 0.192 | -0.12 | -0.18 |

[a] Expanded uncertainties ($k$ = 2): $U(p > 3)/\mathrm{MPa} = 75 \cdot 10^{-6} \cdot \frac{p}{\mathrm{MPa}} + 3.5 \cdot 10^{-3}$; $U(p < 3)/\mathrm{MPa} = 60 \cdot 10^{-6} \cdot \frac{p}{\mathrm{MPa}} + 1.7 \cdot 10^{-3}$; $U(T)$ = 4 mK; $\frac{U(\rho)}{\mathrm{kg \cdot m^{-3}}} = 2.5 \cdot 10^{4} \frac{\chi_{\mathrm{s}}}{\mathrm{m^{3} \cdot kg^{-1}}} + 1.1 \cdot 10^{-4} \cdot \frac{\rho}{\mathrm{kg \cdot m^{-3}}} + 2.3 \cdot 10^{-2}$.

**Table 7.** Experimental ($p$, $\rho_{exp}$, $T$) measurements for the gaseous binary (0.80 $CO_2$ + 0.20 $O_2$) mixture, absolute and relative expanded ($k$ = 2) uncertainty in density, $U(\rho_{exp})$, and relative deviations from the density given by the GERG-2008 [3], $\rho_{GERG}$, and the EOS-CG [6], $\rho_{CG}$, equations of state. Normalized composition of the gaseous mixture is given in Table 2.

| $T$ / K[(a)] | $p$ / MPa[(a)] | $\rho_{exp}$ / kg·m$^{-3}$ | $U(\rho_{exp})$ / kg·m$^{-3}$ | $10^2$ $U(\rho_{exp})/\rho_{exp}$ | $10^2$ $(\rho_{exp}$ - $\rho_{GERG})/\rho_{GERG}$ | $10^2$ $(\rho_{exp}$ - $\rho_{CG})/\rho_{CG}$ |
|---|---|---|---|---|---|---|
| 250 K isotherm | | | | | | |
| 250.048 | 1.951 | 45.821 | 0.034 | 0.075 | 1.18 | 0.33 |
| 250.050 | 1.498 | 33.661 | 0.033 | 0.098 | 0.76 | 0.19 |
| 250.051 | 0.998 | 21.478 | 0.032 | 0.147 | 0.40 | 0.07 |
| 250.051 | 0.499 | 10.322 | 0.030 | 0.293 | 0.09 | -0.06 |
| 260 K isotherm | | | | | | |
| 260.034 | 2.197 | 49.467 | 0.035 | 0.070 | 1.12 | 0.30 |
| 260.035 | 1.521 | 32.407 | 0.033 | 0.101 | 0.64 | 0.16 |
| 260.034 | 0.998 | 20.468 | 0.031 | 0.153 | 0.35 | 0.07 |
| 260.034 | 0.497 | 9.843 | 0.030 | 0.305 | 0.09 | -0.04 |
| 275 K isotherm | | | | | | |
| 275.006 | 4.073 | 97.460 | 0.040 | 0.041 | 2.26 | 0.56 |
| 275.008 | 2.999 | 65.416 | 0.036 | 0.055 | 1.25 | 0.27 |
| 275.007 | 2.006 | 40.820 | 0.033 | 0.081 | 0.68 | 0.13 |
| 275.006 | 0.999 | 19.153 | 0.031 | 0.161 | 0.27 | 0.03 |
| 293.15 K isotherm | | | | | | |
| 293.092 | 7.364 | 201.810 | 0.051 | 0.025 | 4.63 | 0.70 |
| 293.092 | 6.001 | 143.794 | 0.045 | 0.031 | 2.68 | 0.39 |
| 293.093 | 5.000 | 110.844 | 0.041 | 0.037 | 1.82 | 0.24 |
| 293.092 | 4.000 | 82.990 | 0.038 | 0.045 | 1.22 | 0.14 |
| 293.093 | 3.000 | 58.771 | 0.035 | 0.059 | 0.78 | 0.08 |
| 293.093 | 2.000 | 37.241 | 0.032 | 0.087 | 0.44 | 0.03 |
| 293.091 | 0.999 | 17.779 | 0.030 | 0.170 | 0.18 | < 0.01 |
| | | | 300 K isotherm | | | |
| 299.946 | 12.259 | 478.560 | 0.083 | 0.017 | 3.50 | -3.21 |
| 299.948 | 11.048 | 399.019 | 0.074 | 0.018 | 5.93 | -2.02 |
| 299.946 | 10.015 | 326.890 | 0.065 | 0.020 | 6.57 | -0.36 |
| 299.943 | 9.023 | 264.364 | 0.058 | 0.022 | 5.61 | 0.44 |
| 299.948 | 8.003 | 211.441 | 0.052 | 0.025 | 4.16 | 0.50 |
| 299.947 | 7.002 | 169.526 | 0.047 | 0.028 | 3.01 | 0.40 |
| 299.947 | 6.001 | 134.844 | 0.043 | 0.032 | 2.16 | 0.29 |
| 299.949 | 5.000 | 105.303 | 0.040 | 0.038 | 1.54 | 0.20 |
| 299.948 | 4.000 | 79.581 | 0.037 | 0.047 | 1.07 | 0.13 |
| 299.948 | 2.999 | 56.716 | 0.035 | 0.061 | 0.70 | 0.08 |
| 299.949 | 1.999 | 36.107 | 0.032 | 0.089 | 0.41 | 0.04 |
| 299.951 | 0.999 | 17.308 | 0.030 | 0.174 | 0.17 | < 0.01 |
| | | | 325 K isotherm | | | |
| 324.954 | 12.831 | 325.969 | 0.065 | 0.020 | 3.00 | -0.56 |
| 324.955 | 12.035 | 295.650 | 0.061 | 0.021 | 2.93 | -0.40 |
| 324.955 | 11.000 | 257.843 | 0.057 | 0.022 | 2.71 | -0.23 |
| 324.955 | 10.066 | 225.884 | 0.053 | 0.024 | 2.41 | -0.12 |
| 324.955 | 9.022 | 192.910 | 0.050 | 0.026 | 2.07 | -0.03 |
| 324.955 | 8.000 | 163.394 | 0.046 | 0.028 | 1.71 | < 0.01 |
| 324.955 | 7.000 | 136.975 | 0.043 | 0.032 | 1.40 | 0.02 |
| 324.956 | 6.000 | 112.754 | 0.041 | 0.036 | 1.11 | 0.02 |
| 324.956 | 5.000 | 90.434 | 0.038 | 0.042 | 0.86 | 0.02 |
| 324.956 | 4.009 | 69.968 | 0.036 | 0.051 | 0.64 | 0.01 |
| 324.956 | 2.989 | 50.395 | 0.033 | 0.066 | 0.45 | 0.01 |
| 324.956 | 2.000 | 32.665 | 0.031 | 0.096 | 0.28 | < 0.01 |
| 324.956 | 0.999 | 15.825 | 0.030 | 0.187 | 0.13 | < 0.01 |
| | | | 350 K isotherm | | | |
| 349.939 | 13.100 | 260.198 | 0.057 | 0.022 | 1.70 | -0.28 |
| 349.940 | 11.999 | 231.912 | 0.054 | 0.023 | 1.57 | -0.23 |
| 349.938 | 10.996 | 207.061 | 0.051 | 0.025 | 1.44 | -0.18 |
| 349.939 | 9.998 | 183.318 | 0.048 | 0.026 | 1.28 | -0.15 |
| 349.939 | 8.999 | 160.609 | 0.046 | 0.028 | 1.13 | -0.12 |
| 349.939 | 7.998 | 138.940 | 0.043 | 0.031 | 0.97 | -0.09 |
| 349.939 | 6.999 | 118.408 | 0.041 | 0.035 | 0.83 | -0.07 |
| 349.939 | 6.000 | 98.915 | 0.039 | 0.039 | 0.69 | -0.05 |
| 349.939 | 4.999 | 80.345 | 0.037 | 0.045 | 0.56 | -0.03 |
| 349.939 | 4.000 | 62.704 | 0.035 | 0.055 | 0.44 | -0.02 |
| 349.939 | 2.991 | 45.765 | 0.033 | 0.071 | 0.32 | < 0.01 |
| 349.938 | 2.004 | 29.964 | 0.031 | 0.103 | 0.21 | < 0.01 |
| 349.937 | 0.999 | 14.599 | 0.029 | 0.199 | 0.08 | -0.02 |
| | | | 375 K isotherm | | | |
| 374.925 | 12.757 | 212.673 | 0.051 | 0.024 | 1.03 | -0.20 |
| 374.923 | 12.025 | 198.134 | 0.050 | 0.025 | 0.98 | -0.19 |
| 374.923 | 10.978 | 177.772 | 0.047 | 0.027 | 0.90 | -0.16 |
| 374.923 | 9.988 | 159.011 | 0.045 | 0.028 | 0.82 | -0.14 |
| 374.922 | 8.996 | 140.736 | 0.043 | 0.031 | 0.73 | -0.12 |
| 374.924 | 7.998 | 122.913 | 0.041 | 0.033 | 0.64 | -0.10 |
| 374.923 | 6.999 | 105.643 | 0.039 | 0.037 | 0.55 | -0.08 |
| 374.922 | 5.999 | 88.925 | 0.037 | 0.042 | 0.47 | -0.07 |
| 374.924 | 5.005 | 72.873 | 0.035 | 0.049 | 0.38 | -0.05 |
| 374.923 | 4.000 | 57.212 | 0.034 | 0.059 | 0.31 | -0.03 |
| 374.923 | 2.986 | 41.958 | 0.032 | 0.076 | 0.23 | -0.02 |
| 374.922 | 1.999 | 27.603 | 0.030 | 0.110 | 0.14 | -0.02 |
| 374.922 | 0.999 | 13.550 | 0.029 | 0.211 | 0.04 | -0.04 |

[a] Expanded uncertainties ($k$ = 2): $U(p > 3)/\mathrm{MPa} = 75 \cdot 10^{-6} \cdot \frac{p}{\mathrm{MPa}} + 3.5 \cdot 10^{-3}$; $U(p < 3)/\mathrm{MPa} = 60 \cdot 10^{-6} \cdot \frac{p}{\mathrm{MPa}} + 1.7 \cdot 10^{-3}$; $U(T)$ = 4 mK; $\frac{U(\rho)}{\mathrm{kg \cdot m^{-3}}} = 2.5 \cdot 10^{4} \frac{\chi_\mathrm{s}}{\mathrm{m^3 \cdot kg^{-1}}} + 1.1 \cdot 10^{-4} \cdot \frac{\rho}{\mathrm{kg \cdot m^{-3}}} + 2.3 \cdot 10^{-2}$.

**Table 8.** Statistical analysis of the ($p$, $\rho$, $T$) data set processed with the GERG-2008 [3] and EOS-CG [6] for the three ($CO_2$ + $O_2$) mixtures studied in this work, including literature data for comparable mixtures. AAD = absolute average deviation, Bias = average deviation, RMS = root mean square deviation, MaxD = maximum deviation.

| Reference[(a)] | $x(O_2)$ | $N$[(b)] | Covered ranges | | Experimental *vs* GERG-2008 | | | | Experimental *vs* EOS-CG | | | |
|---|---|---|---|---|---|---|---|---|---|---|---|---|
| | | | $T$ / K | $p$ / MPa | AAD / % | Bias / % | RMS / % | MaxD / % | AAD / % | Bias / % | RMS / % | MaxD / % |
| This work | 0.050321 | 45 | 275-375 | 0.5-8 | 0.29 | 0.28 | 0.45 | 1.5 | 0.077 | -0.0022 | 0.11 | 0.41 |
| This work | 0.099856 | 47 | 260-375 | 0.5-9 | 0.69 | 0.68 | 1.1 | 4.4 | 0.15 | 0.054 | 0.26 | 1.2 |
| This work | 0.199907 | 70 | 250-375 | 0.5-13 | 1.3 | 1.3 | 1.9 | 6.6 | 0.22 | -0.052 | 0.50 | 3.2 |
| Gururaja et al. [42] | 0.0-1.0 | 9 | 297-303 | 0.1 | 1.8 | -1.7 | 3.3 | 7.6 | 1.8 | -1.7 | 3.3 | 7.6 |
| Mantovanni et al. [40][(c)] | 0.060700 | 96 | 303-383 | 1-20 | 1.4 | -1.4 | 2.0 | 8.3 | 2.1 | -2.1 | 2.8 | 13 |
| Mantovanni et al. [40][(c)] | 0.129100 | 100 | 303-383 | 1-20 | 2.2 | -2.2 | 2.8 | 13 | 3.5 | -3.5 | 4.1 | 13 |
| Mazzoccoli et al. [41] | 0.044200 | 12 | 273-293 | 1-7 | 2.4 | 2.4 | 2.9 | 6.7 | 1.8 | 1.8 | 2.5 | 6.6 |
| Mazzoccoli et al. [41] | 0.148800 | 18 | 273-293 | 1-7 | 1.2 | 0.4 | 1.8 | 5.2 | 1.9 | -0.89 | 2.5 | 7.2 |
| Al-Siyabi [43] | 0.050000 | 26 | 323-423 | 8-40 | 1.3 | -1.1 | 1.5 | 3.0 | 1.5 | -1.5 | 1.7 | 3.2 |
| Commodore et al. [44] | 0.01246 | 112 | 324-400 | 2-35 | 0.17 | 0.094 | 0.26 | 1.2 | 0.15 | 0.0033 | 0.23 | 1.2 |
| Muirbrook [45] | 0.035-0.4 | 32 | 273.15 | Saturation | 43 | 38 | 75 | 247 | 39 | 34 | 70 | 235 |

[(a)] Only measurements in the vapor and supercritical phase have been considered.

[(b)] Number of experimental points.

[(c)] Used for the EOS-CG development.